\PassOptionsToPackage{dvipsnames}{xcolor}
\documentclass[twocolumn,twocolappendix]{aastex631}
\usepackage{graphicx}
\usepackage[dvipsnames]{xcolor}
\usepackage{todonotes}
\usepackage{multirow}
\usepackage{amsmath}
\usepackage{amssymb}
\usepackage{mathrsfs}

\usepackage{siunitx}
\usepackage{listings}
\usepackage{longtable}
\usepackage{mathtools}

\DeclareRobustCommand{\zzero}{$z = 0$}

\DeclareSIUnit\angstrom{\text {Å}}

\newcommand{\mkold}{\ensuremath{M_K^{\rm 2MASS}}}

\newcommand{\mknew}{\ensuremath{M_K^{\rm CFHT}}}

\newcommand{\mknewmstar}{\ensuremath{M_*\text{--}\mknew}}
\newcommand{\mkmstar}{\ensuremath{M_*\text{--}M_K}}

\newcommand{\mbh}{\ensuremath{M_\mathrm{BH}}}
\newcommand{\mbhmstar}{\ensuremath{\mbh\text{--}M_*}}
\newcommand{\mbhsigma}{\ensuremath{\mbh\text{--}\sigma}}

\newcommand{\msun}{\ensuremath{M_{\odot}}}

\newcommand{\rhobh}{\ensuremath{\rho_{\rm BH}}}
\newcommand{\msunMpc}{\ensuremath{M_{\odot}\,{\rm Mpc}^{-3}}}

\graphicspath{{./figures/}}

\defcitealias{vanderMarelvandenBosch1998}{vdM+vdB 1998}

\begin{document}

\title{Big Galaxies and Big Black Holes: The Massive Ends of the Local Stellar and Black Hole Mass Functions and the Implications for Nanohertz Gravitational Waves}

\correspondingauthor{Emily Liepold}\email{emilyliepold@berkeley.edu}

\author
[0000-0002-7703-7077]
{Emily R. Liepold}
\affiliation{Department of Astronomy, University of California, Berkeley, California 94720, USA.}

\author
[0000-0002-4430-102X]
{Chung-Pei Ma}
\affiliation{Department of Astronomy, University of California, Berkeley, California 94720, USA.}
\affiliation{Department of Physics, University of California, Berkeley, California 94720, USA.}

\begin{abstract}
              
We construct the $z=0$ galaxy stellar mass function (GSMF) by combining the GSMF at stellar masses $M_* \la 10^{11.3} \msun$ from the census study of \citet{lejaetal2020} and the GSMF of massive galaxies at $M_* \ga 10^{11.5} \msun$ from the volume-limited MASSIVE galaxy survey. To obtain a robust estimate of $M_*$ for local massive galaxies, we use MASSIVE galaxies with $M_*$ measured from detailed dynamical modeling or stellar population synthesis modeling (incorporating a bottom-heavy initial mass function) with high-quality spatially-resolved spectroscopy. These two independent sets of $M_*$ agree to within ${\sim}7$\%. Our new $z=0$ GSMF has a higher amplitude at $M_* \ga 10^{11.5} \msun$ than previous studies, alleviating prior concerns of a lack of mass growth in massive galaxies between $z\sim 1$ and 0. We derive a local black hole mass function (BHMF) from this GSMF and the scaling relation of SMBH and galaxy masses.  The inferred abundance of local SMBHs above $\sim 10^{10}\msun$ is consistent with the number of currently known systems. The predicted amplitude of the nanohertz stochastic gravitational wave background is also consistent with the levels reported by Pulsar Timing Array teams.  Our $z = 0$ GSMF therefore leads to concordant results in the high-mass regime of the local galaxy and SMBH populations and the gravitational wave amplitude from merging SMBHs. An exception is our BHMF yields a $z=0$ SMBH mass density that is notably higher than the value estimated from quasars at higher redshifts.

\end{abstract}


\section{Introduction} 

The galaxy stellar mass function (GSMF) specifies the number density of galaxies as a function of their stellar masses at a given redshift. It provides an important characterization of the demographics of galaxies. As galaxies acquire stellar mass via accretion, mergers and star formation, the shape and amplitude of the GSMF change with time. Accurately measured GSMFs at different redshifts therefore inform how galaxies grow and represent a key observational property that must be reproduced in a successful theoretical model or numerical simulation of galaxy formation and evolution. The GSMF can also be used to infer  the black hole mass function (BHMF) because of the strong correlation between the masses of local galaxies and their central supermassive black holes (SMBHs). 

\citet{lejaetal2020} present a recent census of the GSMF spanning the redshift range of $0.2 < z < 3$. Their GSMF is built from $\sim 10^5$ galaxies in the 3D-HST \citep{skeltonetal2014} and COSMOS-2015 \citep{laigleetal2016} catalogs. The stellar masses are determined from photometric properties of the galaxies through the technique of spectral energy distribution (SED) fitting.  As described there, their work is built upon a rich literature of prior measurements of the GSMF from galaxies in different surveys. A major outcome of \citet{lejaetal2020} is a systematically higher number density of galaxies at most stellar masses and redshifts than literature measurements in the preceding decade (e.g., \citealt{liwhite2009, pozzettietal2010, baldryetal2012, santinietal2012, bernardietal2013, muzzinetal2013, moustakasetal2013, tomczaketal2014, mortlocketal2015, davidzonetal2017, wrightetal2018}). The higher amplitude of their GSMF is attributed to differences in the SED-fitting assumptions, where their method produces systematically older stellar ages and higher stellar masses than in previous studies.

Here we present a new analysis of the massive portion of the $z\approx 0$ GSMF at $M_*\ga 3\times 10^{11} M_\odot$, a regime of considerable debate. Several GSMF studies examined in \citet{lejaetal2020} do not sample large enough local volume to have sufficient statistics for determining this part of the GSMF. The exceptions are analyses based on galaxies in the Sloan Digital Sky Survey (SDSS) (e.g., \citealt{liwhite2009, bernardietal2013, moustakasetal2013, dsouzaetal2015}). But the large spatial extents and faint outer envelopes of the local massive galaxies have made it challenging to obtain accurate sky subtraction, surface brightness profile fitting, and total luminosity (and hence stellar mass). \citet{bernardietal2013} find that the number density of high-mass SDSS galaxies is highly dependent on the photometric fitting scheme used to measure the total light. \citet{dsouzaetal2015} estimate a series of flux corrections by stacking images of similar galaxies, finding that the SDSS \texttt{model} magnitudes underrepresent the brightness of the most massive galaxies in their sample by up to 0.4 mags. With these changes,   \citet{dsouzaetal2015} find the number density of $M_*\sim 5\times 10^{11}\msun$ galaxies to be a factor of $\sim 3$ lower than \citet{bernardietal2013} and a factor of $\sim 6$ higher than \citet{liwhite2009}, even though the three analyses all rely on SDSS observations. Furthermore, all of these studies do not explore variations in the stellar initial mass function (IMF) and instead assume a Milky Way IMF.

Reducing the uncertainties in the local GSMF at high masses has far reaching implications. One example is the puzzling lack of net evolution in the massive end of GSMF between $z\sim 1$ and 0 reported in a number of studies (\citealt{moustakasetal2013, bundyetal2017, lejaetal2020}, and references therein). A non-evolving GSMF over this redshift range implies massive galaxies have experienced little mass gain in the past $\sim 8$ billion years, in contrast to the mass growth through galaxy mergers expected since $z\sim 1$.  Another potential puzzle related to the local galaxy and SMBH populations is whether the predicted and detected amplitudes of the stochastic gravitational wave background in the nanohertz regime are consistent with each other. A recent analysis indicates that the amplitudes reported by various Pulsar Timing Array (PTA) teams (NANOGrav, \citealt{NANOGravetal2023}; the European PTA and Indian PTA \citealt{EPTAetal2023}; and the Parkes PTA \citealt{PPTAetal2023}), if originated from merging SMBH binaries, would imply a much higher space density of local massive SMBHs than observed \citep{satopolitoetal2023}. A key link in the models used to predict the gravitational wave amplitude is the local GSMF (or related galaxy velocity dispersion function).
Another consideration is how the GSMF is related to the BHMF inferred from different SMBH populations at various redshifts (e.g., local quiescent vs. high-$z$ active) and the implications for the mass accretion histories of SMBHs (e.g., \citealt{shankaretal2004, kellymerloni2012}).

Here we incorporate results from the MASSIVE survey \citep{maetal2014} in an analysis of the high-mass portion of the local GSMF and the associated BHMF. MASSIVE is a volume-limited multi-wavelength imaging and spectroscopic survey of $\sim 100$ galaxies in the northern sky (declination $\delta > -6^\circ$), targeting all early-type galaxies (ETGs) with $M_* \gtrsim 5\times10^{11}\msun$ to a distance of 108 Mpc. Spatially resolved stellar kinematics of MASSIVE galaxies have been obtained from extensive sets of integral-field spectroscopic observations \citep{vealeetal2017b, vealeetal2017a, Vealeetal2018, eneetal2018, eneetal2019, eneetal2020}.  Hubble Space Telescope (HST) observations of a subset of MASSIVE galaxies and deep $K$-band observations of the majority of the galaxies with WIRCam on the Canada-France-Hawaii Telescope (CFHT) have also been acquired \citep{goullaudetal2018, quennevilleetal2024}. Together, these datasets enable in-depth dynamical modeling to be performed to determine the spatial mass distributions of the stars and dark matter within a galaxy and the mass of the central SMBH.  In addition, deep long-slit spectroscopic observations covering 4000-10300 angstroms have been conducted to measure the spatial gradients of chemical abundances, IMFs, and stellar mass-to-light ratios for a sample of MASSIVE galaxies using stellar population synthesis (SPS) modeling \citep{guetal2022}. These dynamical and SPS measurements of $M_*$ enable us to perform a new assessment of the local high-mass GSMF that is independent of the SED fitting methods used to convert stellar light to mass.

In Section~\ref{sec:Mstar}, we discuss the available sets of dynamical $M_*$ and SPS $M_*$ for MASSIVE galaxies. A list of MASSIVE galaxies with dynamical masses from stellar orbit modeling, stellar Jeans modeling, or gas kinematics is provided in Table~\ref{tab:dynamical_measurements}. We derive a scaling relation between each set of $M_*$ and the $K$-band absolute magnitudes, $M_K$, considering $M_K$ from both 2MASS \citep{skrutskieetal2006} and deeper CFHT observations (Section~\ref{sec:MkMstar}). This relation is then used to infer the stellar mass distribution for the entire MASSIVE sample (Section~\ref{sec:MstarDist}). In Section~\ref{sec:GSMF}, a $z = 0$ GSMF is constructed to reproduce the result of \citet{lejaetal2020} for $M_* \la 10^{11.3}\msun$ and the MASSIVE measurements at $M_* \ga 10^{11.5}\msun$. In Section~\ref{sec:BHMF}, we convolve our GSMF and several GSMFs from the literature with the scaling relation between galaxy mass and SMBH mass to obtain the local BHMF and to estimate the number of massive SMBHs in the local volume. In Section~\ref{sec:implications}, we calculate the amplitude of the gravitational waves in the nanohertz range due to the cosmological distribution of merging SMBHs from each BHMF and compare it with the PTA results. The $z=0$ SMBH mass density from integration of each BHMF is discussed.

\section{Stellar Mass of Local Massive Early-Type Galaxies} \label{sec:Mstar}

When the MASSIVE survey was planned \citep{maetal2014}, the target galaxies were selected based on stellar masses inferred from the 2MASS absolute $K$-band magnitudes and Equation~(2) of \citet{cappellari2013}:
$\log_{10} (M_*/M_\odot) = 10.58 - 0.44 (\mkold + 23)$. 
This empirical formula is a fit to the 2MASS $M_K$ of galaxies in the ATLAS$^{\rm 3D}$ survey \citep{cappellarietal2011} and dynamical masses determined from Jeans modeling (assuming no dark matter). By design, however, there is little overlap in the galaxy mass range probed by the two surveys. The ATLAS$^{\rm 3D}$ survey targets lower-mass early-type galaxies ($M_*\ga 10^{10} M_\odot$) in a smaller volume (to a distance of 42 Mpc); only 6 galaxies in that sample are massive enough to make into the MASSIVE survey. While their \mkmstar\ relation used the best available information at that time, its validity for $M_*$ above $\sim 10^{11.5} \msun$, or $M_K \sim -25$ mag, is untested. There is now a sample of MASSIVE galaxies with dynamical or SPS stellar masses for us to calibrate the relationship between $M_K$ and $M_*$ at high masses. 

\subsection{Stellar Mass from Dynamical Modeling} \label{sec:dynamicalMstar}

In Table~\ref{tab:dynamical_measurements}, we compile a list of MASSIVE galaxies for which dynamical modeling has been performed using detailed spectroscopic measurements of stellar or gas velocities as constraints. Some of these galaxies have been studied with multiple dynamical modeling methods or tracers. In the case of stellar dynamics, eleven galaxies have been modeled with the Schwarzschild orbit method using spatially-resolved stellar kinematics, assuming either an axisymmetric or triaxial potential.  All three major mass components in the galaxies -- SMBH, stars, dark matter -- are included in these orbit models, so we have a direct inference of the dynamical mass of the stellar component of each galaxy. 
The dark matter halo in each case is parameterized by a standard form (e.g., logarithmic or Navarro-Frenk-White). The dark matter fraction within $R_e$ (when reported and when $R_e$ is well determined) ranges from $\sim 15$\% to 40\%.

We note a few caveats to some of the measurements in Table~\ref{tab:dynamical_measurements}. The Jeans stellar dynamical modeling of NGC~5322, NGC~5353, NGC~5557, and NGC~7052 uses a single mass component \citep{Cappellarietal2013a}, so there is not a mass measurement of the stellar component that can be compared fairly with $M_*$ from the orbit method for other galaxies. The CO gas dynamics of NGC~997 and NGC~1684 probe only the inner $\sim 2$ kpc of the galaxies and do not directly constrain the stellar mass beyond this region \citep{dominiaketal2024}. 

\begin{figure*}[ht]
\centering
\includegraphics[width=1.0\textwidth]{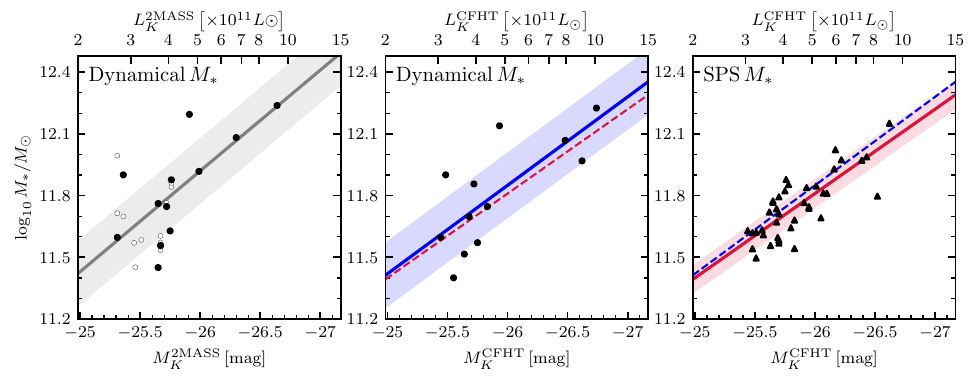}
\caption{
Stellar mass $M_*$ vs. absolute $K$-band magnitude $M_K$ of MASSIVE galaxies with $M_*$ measurements from dynamical modeling (left and middle) or SPS modeling (right). For dynamical $M_*$, two sets of $M_K$ are shown: 2MASS (left) and CFHT (middle). The solid line in each panel represents the respective best-fitting relation (from left to right):
$\log_{10} (M_*/M_\odot) = 11.92-0.49(\mkold + 26)$,
$\log_{10} (M_*/M_\odot) = 11.85-0.43(\mknew + 26)$,
and $\log_{10} (M_*/M_\odot) = 11.81-0.41(\mknew + 26)$;
each shaded band indicates the intrinsic scatter. Filled symbols indicate the measurements used in each fit. Open circles in the left panel mark either multiple $M_*$ measurements of the same galaxy from other methods not used in the fit, or systems excluded from the fit (see Section~\ref{sec:dynamicalMstar} and Table~\ref{tab:dynamical_measurements}). For ease of comparison, the best-fit blue line in the middle panel is shown as dashed blue line in the right panel, and similarly for the red lines, illustrating SPS $M_*$ is on average $\sim 7$\% lower than dynamical $M_*$.}
\label{fig:MstarMk}
\end{figure*}

\subsection{Stellar Mass from Stellar Population Synthesis Modeling} \label{sec:spsMstar}

\citet{guetal2022} present a comprehensive SPS study of 41 MASSIVE galaxies and measurements of their chemical abundances, stellar initial mass functions, and stellar mass-to-light ratios. The high-$S/N$ spectra are obtained from deep spectroscopic observations with LDSS-3 on the Magellan Telescope, covering the spectral range of 4000-10300 angstroms. The mean S/N level (per angstrom) is 120 in the blue and 230 in the red. Since the MASSIVE galaxies are selected to have a declination above $-6^\circ$, only this subset of 41 galaxies are suitable for observations with Magellan.

Table~1 (column 5) of \citet{guetal2022} lists the total $M_*$ from SPS modeling. These values are obtained from the luminosity-weighted $r$-band $M_*/L_r$ within the effective radius, $R_e$, in their SPS model and the total $L_r$ from the Siena Galaxy Atlas \citep{moustakasetal2021, moustakasetal2023} where available or from SDSS. The total galaxy magnitudes from Siena are in SDSS $r$-band and are estimated using the curve of growth method. For consistency with the $K$-band measurements used in this paper, we instead use $M_*/L_K$ (again, luminosity weighted within $R_e$) from the same data and SPS models (Gu et al. in prep) and multiply it by $L_K$ from deep $K$-band CFHT observations (see below) to obtain the SPS-based $M_*$.

We find the two sets of SPS mass-to-light ratios, $M_*/L_r$ and $M_*/L_K$, to be well related by a mean color of $r-K = 2.74$, a value consistent with expectations for a 10 Gyr old stellar population with slightly above solar metallicity \citep{vazdekisetal2012}. When the Siena $L_r$ is shifted by this mean color, we find a residual 0.25 mag difference from the CFHT $L_K$; that is, the Siena luminosities are on average 20\% smaller than those from CFHT. This residual difference could be due to differences in the depths of the two surveys, in the methods used to determine $L$, or uncertainties in SPS modeling.

\subsection{\texorpdfstring{The \mkmstar\ Relation}{The Absolute Magnitude - Stellar-Mass Relation}} \label{sec:MkMstar}

We now examine the correlation between $M_K$ and stellar masses determined by either dynamical or SPS modeling discussed above in the form of
\begin{equation}
   \log_{10} (M_*/M_\odot)
    = \alpha - \beta (M_K + 26)\,.
\label{eq:MkMstar}
\end{equation}   
We consider two sets of measurements of $M_K$ in this analysis: the original 2MASS values, \mkold, from \citet{maetal2014}, and the updated values, \mknew, from the homogeneous deep $K$-band observations with WIRCam on the CFHT Telescope from \citet{quennevilleetal2024}. The \mkold\ values are obtained from the 2MASS apparent $K$-band magnitudes and distances compiled in \citet{maetal2014}. These measurements are useful for a direct comparison with the \mkmstar\ relation in Equation~(2) of \citet{cappellari2013}, which is also based on the 2MASS $K$-band photometry. To improve on the shallow 2MASS photometry, the MASSIVE team has conducted deep CFHT $K$-band observations, reaching 2.5 magnitudes deeper than 2MASS to capture the outer envelopes of these massive galaxies \citep{quennevilleetal2024}. In that work, a non-parametric curve-of-growth method is used to determine the total magnitude of each galaxy; the results therefore do not rely on assuming a particular functional form for the surface brightness profiles, as is frequently done in prior work. The apparent $K$-magnitudes from CFHT are found to be brighter than 2MASS by 0.292 mag on average (see their Figure~2). A complete list of $K^{\rm 2MASS}$ and $K^{\rm CFHT}$ are given in Table~1 of \citet{quennevilleetal2024}.

\subsubsection{\texorpdfstring{Dynamical $M_*$ vs. \mkold}{Dynamical Stellar Masses vs. Absolute Magnitudes from 2MASS}} \label{sec:dyn_mstar_vs_mkold}

We first examine the correlation between \mkold\ and dynamically determined $M_*$ in Section~\ref{sec:dynamicalMstar}, as displayed in Figure~\ref{fig:MstarMk} (left panel). For uniformity, we include only the galaxies in which the stars and dark matter have been reliably modeled as separate components (filled circles). Individual measurement errors are estimated to be 0.1 mag in $M_K$ \citep{quennevilleetal2024, skrutskieetal2006}, and the formal errors on $M_*$ from dynamical modeling tend to be small (e.g., 3.5\% for NGC~1453 in \citealt{quennevilleetal2022} and $<1$\% for NGC~315 in \citealt{boizelleetal2021}). We adopt a uniform error of 0.1 dex on each $M_*$ value to incorporate systematic uncertainties. We find $M_*$ and \mkold\ to be well approximated by Equation~(\ref{eq:MkMstar}) with $\alpha=11.92\pm 0.06$ and $\beta=0.49 \pm 0.16$; the intrinsic scatter is 0.16 dex (grey line and shaded region). This relationship is empirically determined over the magnitude range represented by MASSIVE galaxies. The dominant source of uncertainty in the inferred stellar masses is the intrinsic scatter in the relationship, but if the relationship is extrapolated to substantially fainter magnitudes the uncertainty in the slope $\beta$ will be dominant.

Since both $M_K$ and $M_*$ depend on the distance assumed for a galaxy, we ensure that the same distance is used for the two quantities when deriving the \mkmstar\ relation. The distances used to evaluate \mkold\ were assigned as follows (see Sec. 2.2 of \citealt{maetal2014} for details): for galaxies that belong to groups in the 2MASS Galaxy Redshift Catalog (2MRS; \citealt{huchraetal2012, crooketal2007}), local peculiar velocities were removed and group-corrected redshift distances were assigned; for isolated galaxies with no identifiable groups, redshift distances were assigned based on radial velocities corrected with flow models. We adopt these distances here (column~2 of Table~\ref{tab:dynamical_measurements}). If a different distance was assumed in the relevant dynamical modeling work (column~4), we scale their $M_*$ using the ratio of the distances in columns 2 and 4 in Table~\ref{tab:dynamical_measurements}. 

Our relation between dynamical $M_*$ and \mkold\ has a slightly steeper than the relation of \citet{cappellari2013} and yields a ${\sim}10$\% higher $M_*$ at a fixed \mkold. The actual difference, however, is larger because only a single mass component is modeled in \citet{cappellari2013}, so their reported dynamical masses are upper limits on the true stellar masses. 

\subsubsection{\texorpdfstring{Dynamical $M_*$ vs. \mknew}{Dynamical Stellar Masses vs. Absolute Magnitudes from CFHT}} \label{sec:dyn_mstar_vs_mknew}

For the relation between \mknew\ and dynamically determined $M_*$, we find it well fit by Equation~(\ref{eq:MkMstar}) with $\alpha = 11.85 \pm 0.06, \beta = 0.43 \pm 0.13$, and an intrinsic scatter of 0.16 dex, as indicated by the blue line and shaded region in the middle panel of Figure~\ref{fig:MstarMk}. 

As in the previous section, we have ensured the same distance is used in evaluating $M_*$ and \mknew\ for each galaxy. For uniformity, we adopt the distances used to obtain \mknew\ in \citet{quennevilleetal2024}. For the galaxies with reliably dynamical $M_*$ considered here, the distances have all been measured recently with the surface brightness fluctuation (SBF) method \citep{jensenetal2021, blakesleeetal2021} and are listed in column 3 of Table~\ref{tab:dynamical_measurements}. We scale each $M_*$ from the literature by the ratio of the distances in columns 3 and 4; the adjusted $M_*$ are tabulated in column 8.

While the deeper CFHT observations yield brighter $K$-band magnitudes than 2MASS on average (by 0.292 mag, or a 31\% increase in luminosity),  because the new SBF distances are on average smaller, the \mknew\ in Figure~\ref{fig:MstarMk} is on average 0.12 mag brighter than \mkold.

\subsubsection{\texorpdfstring{Stellar Population Synthesis $M_*$ vs. \mknew}{Stellar Population Synthesis Stellar Mass vs. Absolute Magnitudes from CFHT}}\label{sec:spsMstarMK}

We repeat the analysis in the last subsection using the SPS $M_*$ discussed in Section~\ref{sec:spsMstar}. The SPS $M_*$ and \mknew\ for the 41 galaxies in the \citet{guetal2022} sample are displayed in the right panel of Figure~\ref{fig:MstarMk}. The best fitting relation is given by Equation~(\ref{eq:MkMstar}) with $\alpha = 11.81 \pm 0.02$ and $\beta = 0.41\pm 0.05$; the intrinsic scatter is 0.07 dex (red line and shaded region).   The dynamical $M_*$ and \mknew\ relation is plotted as a dashed blue line for comparison.  The two independent methods used to infer $M_*$ agree well, with the SPS $M_*$ on average $\sim 7$\% smaller than the dynamical $M_*$.

\begin{figure}[t]
\centering
\includegraphics[width=0.95\columnwidth]{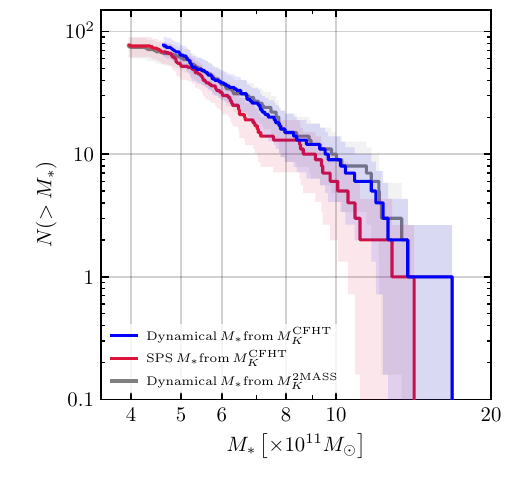}
\caption{
Number of MASSIVE galaxies above stellar mass $M_*$. Each histogram is for $M_*$ from the corresponding \mkmstar\ relation in Figure~\ref{fig:MstarMk}; when available, the dynamical or SPS $M_*$ values are used instead of $M_*$ from the scaling relation. The blue and grey curves are very similar because both are calibrated to MASSIVE galaxies with reliable dynamical $M_*$. The red curve has a lower amplitude because the SPS $M_*$ is on average $\sim 7$\% lower than the dynamical $M_*$. Each shaded band represents the 90\% confidence region.
\label{fig:NaboveMstar}}
\end{figure}

\subsection{Stellar Mass Distribution of the MASSIVE sample} \label{sec:MstarDist}

Having established the \mkmstar\ relation using the subset of MASSIVE galaxies with existing dynamical or SPS $M_*$ in the previous sections, we can now use $M_K$ as a proxy and apply the relation to estimate $M_*$ for the MASSIVE galaxies without these detailed measurements. The distribution of $M_*$ based on each of the three versions of the relation is shown in Figure~\ref{fig:NaboveMstar}.  The two distributions of dynamical-based $M_*$ inferred from \mknew\ (blue curve) and \mkold\ (grey curve) agree well. This is expected because the two \mkmstar\ relations are both calibrated to MASSIVE galaxies with reliable dynamical $M_*$. The distribution of SPS-based $M_*$ (red curve) is lower than that of dynamical-based $M_*$ (blue curve) because the SPS $M_*$ is on average 7\% lower (see Section~\ref{sec:spsMstarMK}). Since the CFHT observations more accurately capture the galaxies' total luminosities than the 2MASS observations, we will use the SPS and dynamical $M_*$ inferred from \mknew\ in the subsequent analysis.

As a final remark, we note that when \mknew\ is used to estimate either the SPS or dynamical $M_*$ above, we need to first homogenize the distance measurements. This step is necessary because even though the galaxies used to calibrate the \mknewmstar\ relation all happen to have SBF distances, SBF distances are available for only $\sim 40$\% of the whole MASSIVE sample.  For this 40\% of galaxies, we find the SBF distances to be $93\% \pm 12\%$ of the 2MRS distances compiled in \citet{maetal2014}.  To avoid introducing systematic biases in MASSIVE galaxies without SBF measurements, we apply this correction factor to the original distances in \citet{maetal2014} and use the SBF-like distances to obtain \mknew. While careful handling of the assumed distance is important for improving the accuracy of each individual stellar mass measurement, the random errors associated with the distance (${\sim}10\%$) are small compared to the intrinsic scatter we found in the \mkmstar\ relations in Section~\ref{sec:MkMstar}.

\section{\texorpdfstring{The Stellar Mass Function at \zzero}{The Local Stellar Mass Function}}\label{sec:GSMF}

In this section, we construct a $z=0$ GSMF that both reproduces the GSMF of \citet{lejaetal2020} below $M_* \sim 10^{11} \msun$ and matches the local measurements based on the volume-limited MASSIVE survey at $M_* \ga 10^{11.5}\msun$  \citep{maetal2014, guetal2022, quennevilleetal2024}. A flat $\Lambda$CDM cosmology with $\Omega_m=0.3$ and $H_0 = 70\,\textrm{km}\,\textrm{s}^{-1}\,\textrm{Mpc}^{-1}$ is assumed.

\subsection{Survey Volume}

To convert the stellar mass distribution of MASSIVE galaxies in Figure~\ref{fig:NaboveMstar} into a mass function, we first estimate the volume surveyed by MASSIVE. To do so, we recall that MASSIVE is primarily a northern-sky survey, targeting all early-type galaxies in the 2MASS catalog satisfying
(i) geometric cuts of distance $D < 108$ Mpc and declination $\delta > -6^\circ$;
(ii) absolute magnitude cuts of $\mkold < -25.3$ mag; and
(iii) extinction cuts of $A_V < 0.6$ in the Extended Source Catalog (XSC; \citealt{jarrettetal2000}) of 2MASS. 

MASSIVE covers 55\% of the sky before the extinction cuts. To estimate the effect of extinction on the survey's sky coverage, we generate a large number of test points uniformly distributed in solid angle with $\delta > -6^\circ$, and then apply the same extinction and reddening relations (\citealt{schlaflyetal2011,fitzpatrick1999}) used in the MASSIVE survey to each point. About 76\% of the points pass the extinction cut of $A_V < 0.6$. MASSIVE therefore effectively covers 41.9\% of the whole sky, or ${\sim}17280\,\textrm{deg}^2$. Folding in the distance cut, we estimate the comoving survey volume of MASSIVE to be $2.05\times10^6\,\textrm{Mpc}^3$. A simple geometric volume of $ \frac{4 \pi}{3} (0.419) (108\,\textrm{Mpc})^3= 2.21\times10^6\,\textrm{Mpc}^3$ would lead to an overestimate of 8\%. The primary source of uncertainty in the survey volume is the assumed distance cut. When computing the GSMF in the next section, we use the differential comoving volume associated with each galaxy's measured distance as a normalization factor rather than assuming a single total survey volume.

For comparison, the sky coverage of the COSMOS-2015 and 3D-HST catalogs used in \citet{lejaetal2020} is ${\sim}2\,\textrm{deg}^2$ and ${\sim}0.25\,\textrm{deg}^2$, respectively. These deep surveys are designed to probe galaxy evolution over a wide redshift range beyond the local volume. The volume covered by these surveys within the distance cut of MASSIVE is tiny in comparison to that of MASSIVE: $237\,\textrm{Mpc}^3$ for COSMOS-2015  and $29.7\,\textrm{Mpc}^3$ for 3D-HST. To ensure a reasonable sample size, \citet{lejaetal2020} limit their GSMF determination to $z > 0.2$ and $z > 0.5$ for the two surveys, respectively.

\subsection{Stellar Mass Function and Schechter Form}\label{sec:gsmf_fitting}

\begin{figure}[t]
\centering
\includegraphics[width=0.88\columnwidth]{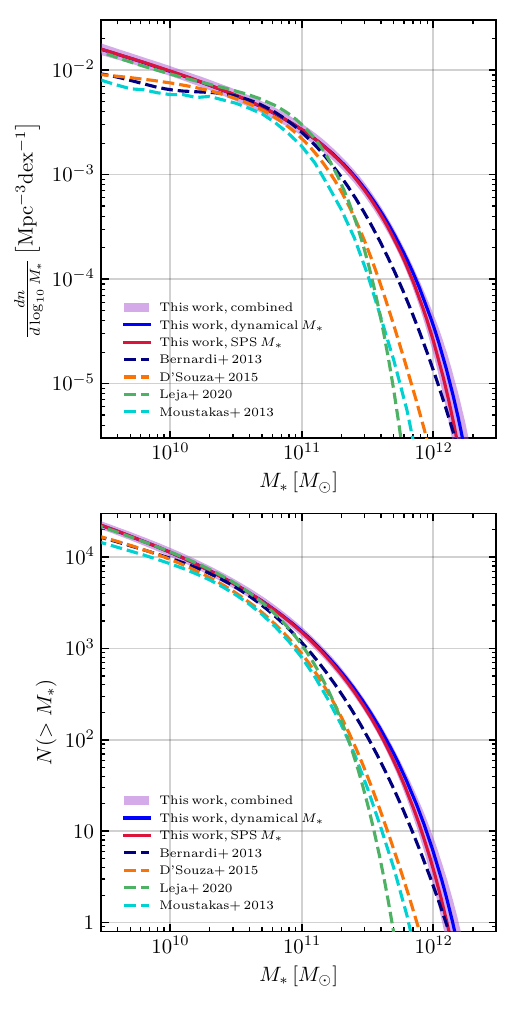}
\caption{
(Top) The $z=0$ galaxy stellar mass function from this paper (violet band; 90\% confidence interval). It is constructed to reproduce the $z=0$ GSMF of \citet{lejaetal2020} (green dashed curve) at $M_* \la 10^{11}\msun$ and to match the stellar mass distribution of MASSIVE galaxies at $M_* \ga 10^{11.5} \msun$. The violet band represents the GSMF from averaging two separate posterior distributions for MASSIVE dynamically inferred $M_*$ (blue curve) and SPS inferred $M_*$ (red curve). Three GSMFs from prior studies of SDSS galaxies are shown for comparison:
\citet{bernardietal2013} (indigo),
\citet{dsouzaetal2015} (orange), 
and \citet{moustakasetal2013} (cyan).
(Bottom) Cumulative galaxy counts in the MASSIVE survey volume estimated from each GSMF in the top panel. The orange, green and cyan curves would yield fewer than 10 galaxies in the MASSIVE survey.
}
\label{fig:GSMF}
\end{figure}

To incorporate the GSMF from \citet{lejaetal2020}, we first make a minor adjustment and extrapolate their $z=0.2$ GSMF (the lowest redshift presented there) to $z=0$. This is achieved using the procedure in their Appendix~B.  This adjustment results in little change at $M_* \ga 10^{11.25} \msun$, about 8\% increase in the amplitude at $M_* \sim 10^{11} \msun$, and about 15\% increase at $M_* \lesssim 10^{10.5} \msun$. The resulting GSMF is plotted as a green dashed curve in the top panel of Figure~\ref{fig:GSMF}. Since the 3D-HST and COSMOS-2015 surveys contain few galaxies above $M_* \sim 10^{11} M_\odot$ in the local volume (Figure~1; \citealt{lejaetal2020}), we include only the  $M_* < 10^{11} M_\odot$ portion of their GSMF in the analysis below.

We find the combined GSMF from MASSIVE and \citet{lejaetal2020} to be well approximated as a sum of two Schechter functions
\begin{equation}\label{eq:gsmf}
 \frac{dn}{d\ln M_*} = \left[ 
 \phi_1 \left( \frac{M_*}{M_s}\right)^{\alpha_1 + 1}
 +  \phi_2 \left( \frac{M_*}{M_s}\right)^{\alpha_2 + 1}  \right]
 e^{-\left(\frac{M_*}{M_s} \right)^\beta}
\end{equation}
where we assume the same scale mass $M_s$ for the two components and $\beta = 1$ as in \citet{lejaetal2020}. We assign the more negative slope to $\alpha_2$ so that the second term is dominant at low masses, while the first term is dominant at high masses. We compute posteriors on the five parameters ($\phi_1, \phi_2, \alpha_1, \alpha_2, M_s)$ in Equation~(\ref{eq:gsmf}) with dynamic nested sampling using \texttt{dynesty} \citep{Speagleetal2020DYNESTY} and list the results in Table~\ref{tab:priorposterior}. This procedure is performed twice, once using MASSIVE dynamically inferred $M_*$ and once using MASSIVE SPS inferred $M_*$, i.e., the blue and red distributions in Figure~\ref{fig:NaboveMstar}, respectively. We use a combined posterior from the average of these two posteriors to obtain our fiducial GSMF. Details on the GSMF fitting procedure and instructions on how to generate the posterior distribution are provided in Appendix~\ref{appendix:phiL_phiH}.

The resulting $z=0$ GSMF is shown as a violet band in the top panel of Figure~3. The median GSMFs from the separate fits to the dynamical and SPS $M_*$ of MASSIVE galaxies are represented by the blue and red curves, respectively. By design, our GSMFs reproduce the $z=0$ GSMF of \citet{lejaetal2020} (green dashed curves) at $M_* \la 10^{11} M_\odot$. For comparison, three additional GSMFs from prior studies of $z\sim 0$ SDSS galaxies are plotted \citep{bernardietal2013, moustakasetal2013, dsouzaetal2015}. The three GSMFs roughly agree at lower masses but start to differ above $M_* \sim 10^{11}\msun$. \citet{bernardietal2013} and \citet{dsouzaetal2015} both adopt improved SDSS magnitude measurements but with different approaches.  The two GSMFs agree well at lower masses but \citet{dsouzaetal2015} is a factor of $\sim 8$ lower at $M_*\sim 10^{12}\msun$.
\citet{moustakasetal2013} do not correct the SDSS magnitudes and instead use stellar masses from the NYU-VAGC catalog \citep{blantonetal2005b}. Their GSMF is a factor of $\sim 3$ lower than \citet{dsouzaetal2015} above $M_* \sim  10^{11.5}\msun$.

Figure~\ref{fig:GSMF} shows that our GSMF (violet band) has a higher amplitude at high masses than the four studies discussed above (dashed curves). It is instructive to compare the number of galaxies expected from each GSMF within the MASSIVE survey volume, which is plotted in the bottom panel of Figure~3.  The GSMF of \citet{bernardietal2013} would predict a factor of $\sim 2$ lower number than the actual galaxy counts in MASSIVE. \citet{lejaetal2020}, \citet{moustakasetal2013} and \citet{dsouzaetal2015} would predict even fewer massive galaxies in this volume, from 0 to $\sim 10$ galaxies with $M_* \ga 5\times 10^{11} M_\odot$, a large discrepancy from the $\sim 100$ galaxies in the MASSIVE survey.

\section{\texorpdfstring{The Black Hole Mass Function at \zzero}{The Local Black Hole Mass Function}}\label{sec:BHMF}

\subsection{BHMFs derived from galaxy distributions}

We use established black hole and galaxy bulge mass scaling relations to convert the GSMF in the previous section into a BHMF. This is achieved by convolution between the GSMF and a probability distribution function associated with the \mbhmstar\ scaling relation:
$\log_{10}(\mbh/\msun) = (8.46\pm0.08) + (1.05\pm0.11) \log_{10} (M_\star/ 10^{11} \msun)$ from \citet{mcconnellma2013}.
This relation assumes log-normal scatter in \mbh\ at a given galaxy mass, so we use the log-normal PDF as the kernel of the convolution. We find that the results are essentially unchanged if the scaling relation of \citet{sagliaetal2016} is used.

\begin{figure}[ht]
\centering
\includegraphics[width=0.99\columnwidth]{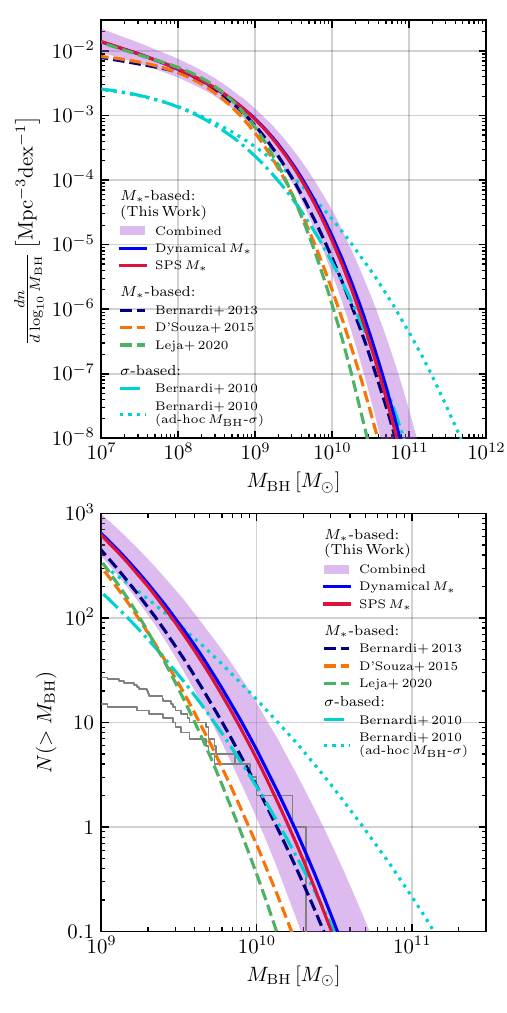}
\caption{
(Top) The $z = 0$ black hole mass function inferred from our GSMF and the \mbhmstar\ scaling relation (violet band; 90\% confidence interval). The dashed curves represent the BHMFs computed from the other GSMFs in Figure~\ref{fig:GSMF}. The two cyan curves compare the BHMFs inferred from galaxy velocity dispersion function instead of GSMF, using the standard \mbhsigma\ relation (dot-dashed) vs. the ad-hoc \mbhsigma\ relation (dotted) from \citet{satopolitoetal2023}.
(Bottom) The SMBH number counts (above mass \mbh) within the MASSIVE survey volume predicted by each BHMF in the top panel. The grey histograms compare the counts of dynamically detected SMBHs within 108 Mpc (top for the entire sphere; bottom for the northern portion surveyed by MASSIVE).
\label{fig:BHMF}}
\end{figure}

The BHMF derived from our GSMF is displayed as a violet band (90\% confidence interval) in the top panel of Figure~\ref{fig:BHMF}. The width of the band arises predominantly from the intrinsic scatter of 0.34 dex in the \mbhmstar\ scaling relation. We find this BHMF to be well approximated by a single Schechter function of the form
\begin{equation}\label{eq:bhmf} 
 \frac{dn}{d\ln \mbh} = 
 \phi \left( \frac{\mbh}{M_s}\right)^{\alpha + 1} e^{-\left( \frac{\mbh}{M_s} \right)^\beta} \,,
\end{equation}
where $\alpha=-1.27\pm 0.02$, $\beta=0.45\pm 0.02$, $\log_{10}(\phi / \rm{Mpc}^{-3})= -2.00\pm 0.07$, and $\log_{10}(M_s/\msun)= 8.09\pm 0.09$.

The BHMFs derived from the GSMFs of \citet{bernardietal2013, dsouzaetal2015} and \citet{lejaetal2020} are plotted as dashed curves for comparison. The uncertainties in these BHMFs (not shown for clarity) are comparable to the violet band because the scatter in the \mbhmstar\ relation is the dominant source of uncertainty. The BHMF inferred from \citet{bernardietal2013} is consistent with the 90\% errorband of our BHMF. The lower amplitudes of the GSMFs of \citet{lejaetal2020} and \citet{dsouzaetal2015} at $M_* \ga 10^{11.3} \msun$ result in lower amplitudes in the BHMFs at $\mbh \ga 10^9 \msun$. 

For further comparison, we include the BHMF (cyan dot-dashed curve) based on the galaxy velocity dispersion (VDF) function and the \mbhsigma\ scaling relation used in \citet{satopolitoetal2023}. The VDF is from \citet{bernardietal2010} and the \mbhsigma\ relation is from \citet{mcconnellma2013}. This $\sigma$-based BHMF overlaps with our BHMF at $\mbh\ga 3\times 10^9 M_\odot$, but at lower \mbh, it predicts a significantly lower number density than any $M_*$-based BHMF shown in Figure~\ref{fig:BHMF}. A similar inconsistency is first reported in \citet{laueretal2007}, which find more SMBHs above $10^9\msun$ when galaxy luminosities instead of $\sigma$ are used as the proxy for \mbh.

As a final comparison, we plot the BHMF (cyan dotted curve) based on the same VDF function but with the hypothetical double power-law \mbhsigma\ relation proposed in \citet{satopolitoetal2023}. The authors introduce a much steeper second power law in this ad-hoc \mbhsigma\ relation in order to match the amplitude of the stochastic gravitational waves background observed by PTA teams.  As pointed out in that work, this BHMF would predict significantly more SMBHs above $\mbh \sim 10^9 M_\odot$ than if the standard single power-law \mbhsigma\ relation were used (cyan dot-dashed curve). This BHMF also far exceeds our $M_*$-based BHMF at $\mbh \ga 10^{10.3} M_\odot$ (violet band). We will discuss the implications below.

\subsection{How many big black holes are there?}

To estimate the expected number of SMBHs above mass $\mbh$ in the local universe, we integrate each $z=0$ BHMFs discussed above and multiply by the MASSIVE survey volume. The resulting number distributions of $\mbh > 10^9 M_\odot$ SMBHs are shown in the bottom panel of Figure~\ref{fig:BHMF}. To compare with known SMBHs, we compile a list of SMBHs with dynamically determined masses above $10^9 M_\odot$ out to the MASSIVE survey distance (108 Mpc) and plot the mass distribution as grey histograms (top for the entire sphere; bottom for the northern portion surveyed by MASSIVE). This list is based on the samples from \citet{mcconnelletal2013} and \citet{sagliaetal2016} and new SMBHs in this mass range and volume reported since those compilations \citep{walshetal2015, thomasetal2016, walshetal2016, walshetal2017, boizelleetal2021, quennevilleetal2022, pilawaetal2022, liepoldetal2023, denicolaetal2024, dominiaketal2024, mehrganetal2024}.

The BHMF based on the GSMF proposed in this paper (violet band) predicts 1 to 14 (90\% confidence interval) SMBHs with $\mbh \ga 10^{10} \msun$, and between 0.1 to 2.2 at $\mbh \ga 2\times 10^{10} \msun$, consistent with the known number of SMBHs in this mass range. In comparison, the cyan dotted curve proposed in \citet{satopolitoetal2023} predicts many more ultra-massive SMBHs than observed.

Another noteworthy feature of the bottom panel of Figure~\ref{fig:BHMF} is all curves predict many more $\mbh \sim 10^9 M_\odot$ SMBHs than currently known. This is a reflection of the incomplete census of local SMBHs. Direct dynamical inference of a SMBH and a robust measurement of its mass require time-consuming observational and modeling efforts. Among the most massive $\sim 100$ local galaxies targeted by the MASSIVE survey, only $\sim 15$ SMBHs have been inferred from detailed dynamical methods thus far (Table~\ref{tab:dynamical_measurements}). Ongoing efforts are expected to increase this sample.

\section{Implications}\label{sec:implications}

\subsection{Amplitude of Stochastic Gravitational Waves}\label{sec:strain}

The cosmological distribution of merging SMBH binaries is expected to produce a stochastic gravitational wave background with a characteristic spectrum and amplitude that depend on the properties of the mergers. Consider a SMBH binary at redshift $z$ with a mass ratio $q = M_2 / M_1 \le 1$, total mass $M = M_1 + M_2$, and chirp mass $\mathcal{M}$ where $\mathcal{M}^{5/3} = M^{5/3}q / (1 + q)^2$. The distribution of these binaries is represented by the mass function $d^3n(M, q, z)/dM\,dq\,dz$, which defines the differential number density of binaries with $M, q$, and $z$. For SMBH binaries on circular orbits where the orbital decay is purely driven by gravitational radiation, the characteristic amplitude of the gravitational waves at frequency $f$ (over interval $d\ln f$) can be written as \citep{phinney2001}
\begin{eqnarray}
\label{eq:strain0}
 h_c^2(f) &&=  \frac{4 \pi}{3 c^2} \frac{1}{(\pi f)^{4/3}} \\
&& \times\int dM\, dq\, dz \frac{d^3n}{dM\,dq\,dz} 
  \frac{q (GM)^{5/3}}{(1 + q)^2}\frac{1}{(1 + z)^{1/3}} \nonumber \,.
\end{eqnarray}

\begin{figure*}[ht]
\centering
\includegraphics[width=0.85\textwidth]{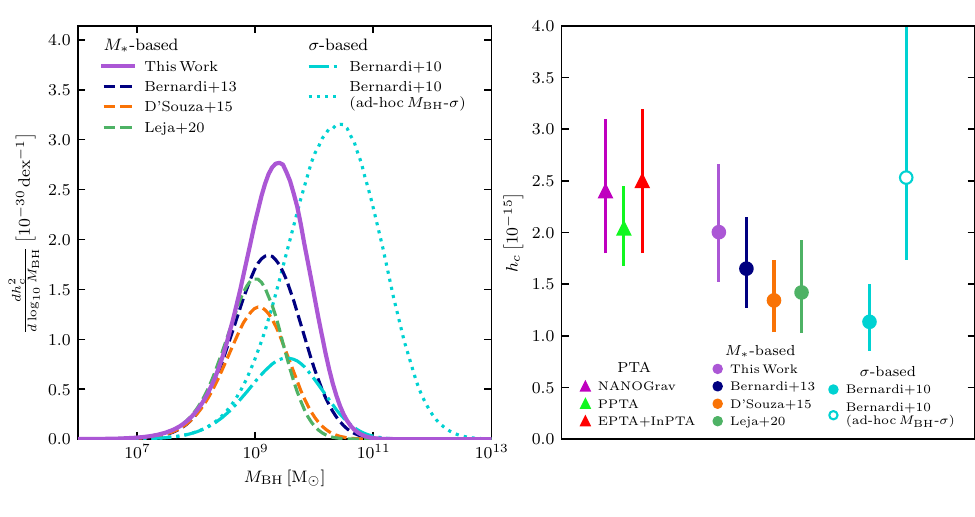}
\caption{
Comparison of the characteristic amplitude of gravitational waves at $f = 1\, {\rm yr}^{-1}$, $h_c$, reported by the PTA teams and predicted by BHMFs inferred from galaxy distribution functions. The right panel shows $h_c$ and the left panel shows the differential contribution to $h^2_c$, $d h^2_c/d\log_{10}\mbh$, as a function of \mbh. The predicted amplitudes are computed from Equation~(\ref{eq:strain}) using the BHMFs in Figure~\ref{fig:BHMF}. All errorbars denote 90\% confidence intervals. The published PPTA value quotes a 68\% interval; we enlarge  it by a factor of 1.64 to approximate the 90\% interval. The individual value of $h_c$ (in units of $10^{-15}$) is (from left to right):
$(2.4^{+0.7}_{-0.6}), 
(2.04^{+0.41}_{-0.36}),
(2.5\pm 0.7)$ from PTAs; 
$(2.0^{+0.7}_{-0.5}), 
(1.7^{+0.5}_{-0.4}),  (1.3^{+0.4}_{-0.3}),  (1.4^{+0.5}_{-0.4})$ from the four $M_*$-based GSMFs; and $(1.1^{+0.4}_{-0.3}), (2.5^{+1.5}_{-0.8})$ from the $\sigma$-based GSMFs.
}
\label{fig:strain}
\end{figure*}

\begin{figure*}
\centering
\includegraphics[width=0.85\textwidth]{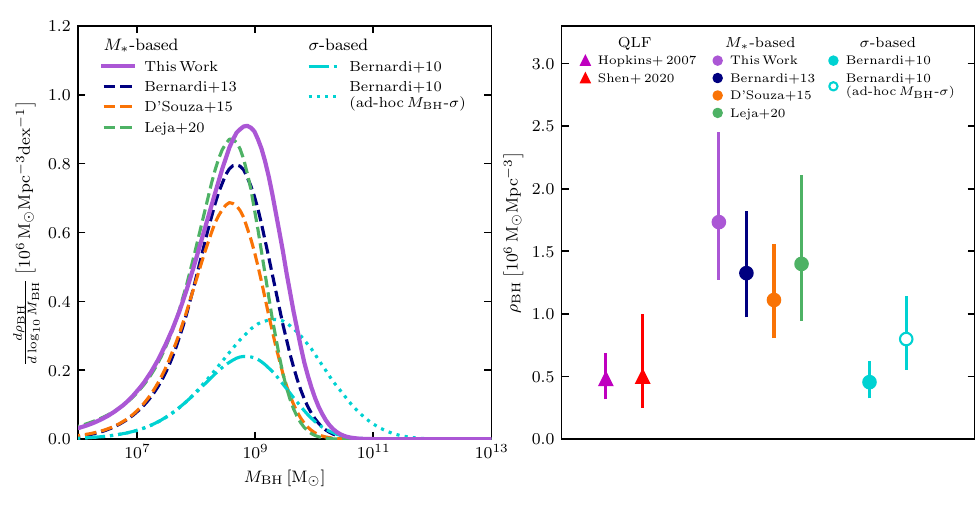}
\caption{
Comparison of the local black hole mass density, \rhobh, given by each BHMF in Figure~\ref{fig:BHMF} and inferred from the quasar luminosity function. The right panel shows \rhobh\ and the left panel shows the differential contribution to \rhobh, $d \rhobh/d\log_{10}\mbh$, as a function of \mbh. All errorbars denote 90\% confidence intervals. \citet{hopkinsetal2007} quotes a 68\% interval; we enlarge it by a factor of 1.64 to approximate the 90\% interval. }
\label{fig:rhobh}
\end{figure*}

\citet{satopolitoetal2023} further assume that the distributions in $M, q$ and $z$ are separable and replace $d^3 n/dM\,dq\,dz$ with $p_z(z)\, p_q(q)\, dn/dM$, where $p_z(z)$ and $p_q(q)$ are the normalized distributions of redshifts and mass ratios, respectively. They also assume that each present-day SMBH has experienced a single merger in its lifetime and identify $dn/dM$ as the local BHMF with $M=\mbh$. The strain is then simplified to
\begin{eqnarray}
\label{eq:strain}
    h_c^2(f) && = 
    \frac{4 \pi}{3 c^2} \frac{1}{(\pi f)^{4/3}} \,
 \langle q/(1 + q)^2 \rangle
\langle (1+z)^{-1/3} \rangle\nonumber \\
&&  \qquad \times \int dM \frac{dn}{dM} (GM)^{5/3}   \\
&& =  1.18\times 10^{-30} \left(\frac{{\rm yr}^{-1}}{f}\right)^{4/3}
\langle q/(1 + q)^2\rangle
\langle (1 + z)^{-1/3} \rangle \nonumber\\
&& \qquad \times \int d M  \left(\frac{M}{10^9\msun}\right)^{5/3} \frac{d}{dM} \left(\frac{n}{10^{-4}{\rm Mpc}^{-3}}\right) \nonumber \,
\end{eqnarray} 
where ${\langle q/(1 + q)^2\rangle = \int dq\,p_q q / (1 + q)^2}$ and ${\langle (1 + z)^{-1/3}\rangle = \int dz\,p_z (1 + z)^{-1/3}}$. \citet{satopolitoetal2023} adopt 
$p_z(z) \propto z^\gamma e^{-z /z_*}$ with $\gamma = 0.5$ and $z_* = 0.3$ and 
$p_q(q) \propto q^\delta$ for $0.1 < q < 1$ with $\delta = -1$. The two distributions peak at  $z \sim 0.3$ and $q= 0.1$, and  $\langle q/(1 + q)^2\rangle = 0.178$ and  $\langle (1 + z)^{-1/3}\rangle = 0.894$. In comparison, the SMBH population synthesis modeling in \citet{Nanogravetal2023_L37} finds the source contribution to $h_c^2$ to be peaked near $z \sim 1$ and $q \sim 1$ (see solid curves in their Figure~12). We find their distributions well approximated by
\begin{equation}
p_z(z) \propto z^\gamma e^{-(z /z_*)^\beta}\,, \quad
p_q(q) \propto q^2 \,,
\label{eq:pzpq}
\end{equation}
and $\gamma = 1.0$, $z_* = 0.5$, and $\beta = 2$. Using these distributions yields $\langle q/(1 + q)^2\rangle = 0.238$ and  $\langle (1 + z)^{-1/3}\rangle = 0.890$, and raises the inferred $h_c$ by about 14\% compared to those of \citet{satopolitoetal2023}. We adopt Equation~(\ref{eq:pzpq}) for the following calculations.

The amplitude $h_c$ (at $f=1 {\rm yr}^{-1}$) computed from Equations~(\ref{eq:strain}) and (\ref{eq:pzpq}) using each of the BHMFs discussed in Section~\ref{sec:BHMF} is displayed in the right panel of Figure~\ref{fig:strain}. 
The first four points show that the 90\% confidence intervals of $h_c$ from the new BHMFs presented in this work (violet circle)
overlap substantially with the PTA results (for a fixed spectral index of $\alpha = -2/3$).  Therefore, unlike \citet{satopolitoetal2023}, we do not find tension between $h_c$ predicted by our BHMF and reported by the PTAs.  Our calculation also shows that 
$h_c$ inferred from earlier GSMFs
are somewhat lower than the PTA levels, but the 90\% confidence intervals largely overlap.  However, when the BHMF inferred from galaxy $\sigma$ distribution is used, 
we obtain a low $h_c$ (filled cyan circle), consistent with the finding of \citet{satopolitoetal2023}. We have also verified that their ad hoc broken power law \mbhsigma\ relation with $\mbh \sim \sigma^{10.5}$ at the high mass end indeed boosts $h_c$ to be compatible with PTA values (open cyan circle). 

The origin of the differing $h_c$ is illustrated in the left panel of Figure~\ref{fig:strain}, where we plot the differential contribution to $h^2_c$, $d h^2_c/d\log_{10}\mbh$, as a function of \mbh\ for each model. 
For all the calculations based on GSMFs, the contributions to $h_c$ peak at $\mbh \sim (1-3)\times 10^9 \msun$.  The main difference among the various GSMFs is the height of this peak. 
The $\sigma$-based BHMF has a different shape and a lower amplitude at $\mbh \la 5\times 10^9\msun$ than all the $M_*$-based BHMFs (top panel of Figure~\ref{fig:BHMF}), 
leading to a lower peak in the mass range $\mbh \sim (1-5)\times  10^9\msun$ that contributes most to $h_c$.
In comparison, the ad hoc model of \citet{satopolitoetal2023}, which is designed to match PTA results, moves the peak of the $h_c$ contribution to $\mbh \sim 3\times10^{10}\msun$ (dotted cyan curve), and hence their requirement of a population of $3\times 10^{10} \msun$ SMBHs. By contrast, our GSMF and BHMF are able to match the PTA levels by having a higher amplitude of GSMF at $M_*\sim 10^{12}\msun$ and a higher amplitude of BHMF at $\mbh \sim 2\times 10^9\msun$ without the need for a population of unobserved ultra-massive SMBHs in the local volume.

As a final note, we perform a quick estimate of $h_c$ by approximating the BHMF as a broken power law over the masses where the majority of the contribution to $h_c$ lies. Our BHMF is roughly $dn/d\log_{10}\mbh \propto \mbh^{-1.1}$ for $10^{7.5} \msun < \mbh < 10^{9.5} \msun$ and $\propto \mbh^{-2.5}$ for $10^{9.5}\msun < \mbh < 10^{10.5} \msun$, where the amplitude at $\mbh = 10^{9.5}\msun$ is $\sim 2\times 10^{-4}\,{\rm Mpc}^{-3}$. Using this approximation and the simplifying assumption of $q=1$ and $z=1$ in Equation~(\ref{eq:strain}), we find $h_c \sim 1.9\times 10^{-15}$, which is within $\sim 5$\% of $h_c = 2.0\times 10^{-15}$ from our more careful calculation.

\subsection{Local Black Hole Mass Density}\label{sec:rhobh}

Various approaches have been taken to estimate the local mass density of SMBHs from properties of galaxies (see review by \citealt{kellymerloni2012} and references therein). One common method uses properties of quasars and links SMBH mass growth to the quasar luminosity function (QLF) via some variation on the So\l{}tan argument \citep{soltan1982}. The estimates have typically yielded a SMBH mass density in the range of $\rhobh = (2-6.5)\times 10^5 \msunMpc$
if the quasar mass-to-energy conversion efficiency is assumed to be $\epsilon\sim 0.1$ (e.g., \citealt{saluccietal1999, yutremaine2002, shankaretal2004, marconietal2004, hopkinsetal2007}). A recent update to \citet{hopkinsetal2007} using newer quasar SED model, bolometric and extinction corrections, and binned estimations of the QLF finds a factor of $\sim 4$ uncertainty: $\rhobh = (2.5-10)\times 10^5 \msunMpc$ \citep{shenetal2020}.

Another approach to estimate the local $\rho_{\rm BH}$ takes into account the entire local galaxy population (as we have done in this paper). The SMBH mass density is obtained by convolving the distribution function of a galaxy property (e.g., velocity dispersion $\sigma$, luminosity $L$, stellar mass $M_*$) 
with a scaling relation between that galaxy property and \mbh. \citet{yutremaine2002} find the $\sigma$-based density to be $\rho_{\textrm{BH}} = (2.9\pm 0.5)\times 10^5\msun$ Mpc$^{-3}$ and the $L$-based density to be $\rho_{\textrm{BH}} = 6.3\times 10^5\msun$ Mpc$^{-3}$ (after scaling their results to $h=0.7$). While \citet{yutremaine2002} favor galaxy $\sigma$ as the \mbh\ predictor, a subsequent study by \citet{laueretal2007} advocates galaxy $L$ as the \mbh\ predictor and find $\rho_{\textrm{BH}} = 4.4\times 10^5\msun$ Mpc$^{-3}$ with similar calculations but updated data. They report a lower value of $\rho_{\textrm{BH}} = 1.6\times 10^5\msun$ Mpc$^{-3}$ when $\sigma$ is used. A compilation of results from various galaxy indicators and data obtains $\rhobh = (3.2-5.4)\times 10^5\msun$ Mpc$^{-3}$  \citep{shankaretal2009}.

The right panel of Figure~\ref{fig:rhobh} shows our estimates of \rhobh\ from integration of the BHMFs presented in Figure~\ref{fig:BHMF}; the left panel shows the corresponding differential contribution $d\rhobh/d\log_{10}{\mbh}$ as a function of \mbh. The $M_*$-based values for \rhobh\ are all above $\sim 10^6 \msunMpc$, with $\rhobh=(1.8^{+0.8}_{-0.5})\times 10^6\msunMpc$ for our BHMF, and the differential contributions to $\rhobh$ all peak at $\mbh\sim 10^{8.5}-10^9 \msun$. Our $\sigma$-based $\rhobh=(4.55^{+1.71}_{-1.28})\times 10^5 \msunMpc$ is lower by a factor of 2 or more (filled cyan circle), similar to the difference between the $\sigma$ and $L$ predictors discussed above. Our $\sigma$-based value is about 50\% higher than the $\sigma$-based value of \citet{yutremaine2002} largely because of the steeper slope of the \mbhsigma\ relation in \citet{mcconnellma2013} than \citet{tremaineetal2002}.  A similarly steep \mbhsigma\ relation is reported in \citet{sagliaetal2016}, which we find to give essentially the same \rhobh\ as ours.

A notable trend in Figure~\ref{fig:rhobh} is all $\rhobh$ inferred from BHMFs based on galaxy stellar masses are significantly higher than the values based on QLF discussed above (two examples shown as triangles). Since the mass acquired during the bright quasar phases scales roughly inversely with the assumed radiative efficiency $\epsilon$, one way to match the high values of local $\rhobh$ found in this paper is to have $\epsilon \la 0.03$. \citet{shankaretal2013} examine but reject a model in which $\epsilon$ decreases from $\sim 0.14$ at high redshift to $\sim 0.004$ at present day because it suggests a larger relic BHMF than their data at most mass scales. Alternatively, raising the obscuration fraction of AGNs can help boost  the inferred local \rhobh. 

\section{Conclusions}

We have presented a new $z=0$ GSMF and calculated the inferred local BHMF, number of massive SMBHs, SMBH mass density, and the amplitude of the stochastic gravitational wave background probed by PTA teams. Our GSMF is constructed to match the observed distribution of stellar masses of MASSIVE galaxies at $M_* \gtrsim 10^{11.5}\msun$. We have incorporated two sets of $M_*$ measurements from dynamical modeling and SPS modeling and used them to calibrate the $\mkmstar$ relation for the whole MASSIVE sample. These state-of-the-art $M_*$ measurements are obtained from high-$S/N$ and spatially-resolved stellar spectroscopic data and detailed modeling. Encouragingly, we find the average $M_*$ from these two independent methods to differ by only $\sim 7$\% (Figure~\ref{fig:MstarMk}). 

While prior $z\sim 0$ GSMFs differ significantly from one another at $M_* \ga 10^{11.3}\msun$, our GSMF has the highest number density of massive galaxies (Figure~\ref{fig:GSMF}). These prior studies measure $M_*$ from photometric data using a variety of SED- and SPS-fitting methods assuming a Chabrier-like or Kroupa-like IMF. The spectroscopic study of \citet{guetal2022}, however, finds the IMF of MASSIVE galaxies to be steeper than the Milky Way IMF with an average mismatch parameter of $\alpha_{IMF}\equiv (M/L)/(M/L)_{\rm MW} = 1.84$. While a re-analysis of prior work with a bottom-heavy IMF is needed to understand better the origin of the differing GSMFs, we note that a simple shift in $M_*$ rightward by a factor of $\alpha_{IMF}$ in the high-mass portion of the \citet{bernardietal2013} and \citet{ dsouzaetal2015} curves in Figure~\ref{fig:GSMF} would bring them close to our GSMF. 

A factor of 2.5 to 3 increase in $M_*$, however, would be needed for the high-mass end of the $z=0$ \citet{lejaetal2020} GSMF to match ours. Thus, even if we assume the same $\alpha_{IMF}$ correction factor at $z\sim 1$ and shift their $z\sim 1$ GSMF accordingly, there would still be notable mass growths in massive galaxies between their $z\sim 1$ GSMF and our $z\sim 0$ GSMF. Our GSMF therefore largely resolves the inconsistency between a lack of evolution in prior GSMFs between $z\sim 1$ and $z=0$ and the mass growth over these 8 billion years expected from galaxy formation models and simulations.

The BHMF derived from our GSMF  (Figure~\ref{fig:BHMF}) predicts $\sim 1-10$ SMBHs with $\mbh\ga 10^{10} \msun$ in the MASSIVE survey volume, fully consistent with the current known SMBH population at the highest masses. The characteristic amplitude of nanohertz gravitational waves due to SMBH binary mergers inferred from our BHMF is also consistent with the levels reported by the PTAs (Figure~\ref{fig:strain}). However, we find a substantially smaller $h_c$ when galaxy velocity dispersion is used as a proxy for \mbh. It has been noted that galaxy velocity dispersions tend to underpredict \mbh\ in massive galaxies compared with galaxy luminosities or stellar masses (e.g., \citealt{laueretal2007}). A plausible explanation is gas-poor mergers of elliptical galaxies that are primarily responsible for the formation of local massive galaxies can easily increase $L$ and $M_*$ while changing $\sigma$ slowly (e.g., \citealt{boylankolchinetal2006}). Galaxy $\sigma$ is therefore likely to be a less robust indicator of \mbh\ for very massive galaxies in the local universe.

While our new $z=0$ GSMF leads to a concordant picture in the high-mass end of the local galaxy and SMBH populations and the gravitational wave background from merging SMBHs, one intriguing difference is the local SMBH mass densities inferred from our and other GSMFs are notably higher than the SMBH mass density estimated from quasars (Figure~\ref{fig:rhobh}). We leave this topic to future investigations. 

\section*{Acknowledgements}

We thank Joel Leja for stimulating discussions. We acknowledge support of NSF AST-2206307, NSF AST-2307719, the Heising-Simons Foundation and the Miller Institute for Basic Research in Science.

\appendix

\section{Dynamical Masses of MASSIVE Galaxies}\label{sec:appendix:MASSIVE}
\setcounter{table}{0}
\renewcommand{\thetable}{\ref*{sec:appendix:MASSIVE}\arabic{table}}

Table~\ref{tab:dynamical_measurements} lists galaxies in the MASSIVE survey with masses that have been determined from dynamical modeling.

\begingroup\scriptsize
\setlength{\LTcapwidth}{\textwidth}
\begin{longtable*}[c]{lcccccccll}

\multicolumn{10}{c}{{\bf Table~\ref{tab:dynamical_measurements}}}\\
\multicolumn{10}{c}{Dynamical masses of MASSIVE galaxies}\\

\hline\hline
Name & $D_{\textrm{M14}}$ & $D_{\textrm{SBF}}$  & $D_{\textrm{literature}}$ & \mkold\ & \mknew\ & \mbh  & $M_*^{dyn}$ & Method & Reference  \\
      & (Mpc) & (Mpc) & (Mpc) & (mag) & (mag) & ($10^9 \msun$) &
     ($10^{11} \msun$) & & \\
        (1) & (2) & (3) & (4) & (5) & (6) & (7) & (8) & (9) & (10) \\ \hline
 \endfirsthead

\hline\hline

\endhead
\hline
\endfoot
\hline
\caption{
Column 1: MASSIVE galaxy name. 
Column 2: distance from MASSIVE survey paper \citet{maetal2014},
which adopt SBF distance if available and group-corrected flow velocities otherwise.
Column 3: updated distance from \citet{quennevilleetal2024}, all by the SBF method except NGC~997, for which the non-SBF distance is corrected as described in Section~\ref{sec:MstarDist}.
Column 4: distance assumed in the dynamical modeling literature.
Column 5: extinction-corrected total absolute $K$-band magnitude derived from 2MASS XSC apparent $K$-band magnitude (parameter k\_m\_ext; column 5 of Table 3 in \citealt{maetal2014}) and distance in Column 3.
Column 6: absolute $K$-band magnitude derived from CFHT apparent $K$-band magnitude (column 6 of Table 1 in \citealt{quennevilleetal2024}) and distance in Column 3.
Column 7: black hole mass from dynamical modeling, corrected to distance in Column 3.
Column 8: total stellar mass from 
dynamical modeling, corrected to distance in Column 3. Symbol $\dagger$ indicates total dynamical mass when mass of the stellar component is unavailable. The 11 mass measurements used in Section~\ref{sec:Mstar} are indicated with asterisks ($*$).
Column 9: method used for dynamical modeling. ``Tri." for Trixial, 
``Axi." for Axisymmetric, ``Sph.'' for Spherical. 
Column 10: dynamical modeling reference.
\label{tab:dynamical_measurements}
}
\endlastfoot

NGC~57   &  $76.3$    &  $66.9$    &  $66.8$    &  $-25.75$  &  $-25.75$      &      &  ${  3.73^*}$     &  {\scriptsize Tri. stellar orbit }  &  {\scriptsize Pilawa et al. in prep. } \\
NGC~315   &  $70.3$    &  $68.1$    &  $70.0$    &  $-26.30$  &  $-26.48$      &  $2.02$    &  ${  11.70^*}$    &  {\scriptsize CO gas}  &  {\scriptsize \citet{boizelleetal2021} } \\
NGC~708   &  $69.0$    &  $61.5$    &  $68.5$    &  $-25.65$  &  $-25.55$      &  $8.98$    &  ${ 2.51^*}$            &  {\scriptsize Tri. stellar orbit }  &  {\scriptsize \citet{denicolaetal2024} } \\
NGC~997   &  $90.4$    &  $83.7^*$    &  $90.4$    &  $-25.40$  &  $-25.68$      &  $3.28$    &                   &  {\scriptsize CO gas }  &  {\scriptsize \citet{dominiaketal2024} } \\
NGC~1453   &  $56.4$    &  $51.2$    &  $51.0$    &  $-25.67$  &  $-25.64$      &  $2.91$    &  ${  3.28^*}$     &  {\scriptsize Tri. stellar orbit }  &  {\scriptsize \citet{quennevilleetal2022} } \\
          &  $56.4$    &  $51.2$    &  $51.0$    &  $-25.67$  &  $-25.64$      &  $2.91$    &  $3.39$            &  {\scriptsize Axi. stellar orbit }  &  {\scriptsize \citet{liepoldetal2020} } \\
          &  $56.4$    &  $51.2$    &  $56.4$    &  $-25.67$  &  $-25.64$      &  $2.99$    &  $3.65$            &  {\scriptsize Axi. stellar Jeans }  &  {\scriptsize \citet{eneetal2019} }  \\
NGC~1600   &  $63.8$    &  $71.7$    &  $64.0$    &  $-25.99$  &  $-26.62$      &  $19.05$   &  ${  9.30^*}$     &  {\scriptsize Axi. stellar orbit }  &  {\scriptsize \citet{thomasetal2016} } \\
NGC~1684   &  $63.5$    &  $62.8$    &  $62.8$    &  $-25.34$  &  $-25.70$      &  $1.40$    &                   &  {\scriptsize CO gas }  &  {\scriptsize \citet{dominiaketal2024} } \\
NGC~2693   &  $74.4$    &  $71.0$    &  $71.0$    &  $-25.76$  &  $-25.72$      &  $1.70$    &  ${  7.19^*}$     &  {\scriptsize Tri. stellar orbit }  &  {\scriptsize \citet{pilawaetal2022} } \\
          &  $74.4$    &  $71.0$    &  $71.0$    &  $-25.76$  &  $-25.72$      &  $2.40$    &  $6.94$            &  {\scriptsize Axi. stellar orbit }  &  {\scriptsize \citet{pilawaetal2022} } \\
          &  $74.4$    &  $71.0$    &  $71.0$    &  $-25.76$  &  $-25.72$      &  $2.90$    &  $6.64$            &  {\scriptsize Axi. stellar Jeans }  &  {\scriptsize \citet{pilawaetal2022} } \\
NGC~3842   &  $99.4$    &  $87.5$    &  $98.4$    &  $-25.91$  &  $-25.93$      &  $8.63$    &  ${  13.78^*}$    &  {\scriptsize Axi. stellar orbit }  &  {\scriptsize \citet{mcconnelletal2011a} } \\
NGC~4472   &  $16.7$    &  $16.7$    &  $17.1$    &  $-25.72$  &  $-25.83$      &  $2.44$    &  ${  5.58^*}$     &  {\scriptsize Axi. stellar orbit }  &  {\scriptsize \citet{ruslietal2013} } \\
          &  $16.7$    &  $16.7$    &  $17.1$    &  $-25.72$  &  $-25.83$      &            &  $5.82^\dagger$            &  {\scriptsize Axi. stellar Jeans }  &  {\scriptsize \citet{cappellarietal2013} } \\
M87
&  $16.7$    &  $16.7$    &  $16.8$    &  $-25.31$  &  $-25.44$      &  $5.34$    &  ${  3.95^*}$     &  {\scriptsize Tri. stellar orbit }  &  {\scriptsize \citet{liepoldetal2023} } \\
          &  $16.7$    &  $16.7$    &  $17.9$    &  $-25.31$  &  $-25.44$      &  $6.16$    &  $9.86$            &  {\scriptsize Axi. stellar orbit }  &  {\scriptsize \citet{Gebhardtetal2011} } \\
          &  $16.7$    &  $16.7$    &  $17.9$    &  $-25.31$  &  $-25.44$      &  $3.27$    &                    &  {\scriptsize Ionized gas  }  &  {\scriptsize \citet{walshetal2013} } \\
          &  $16.7$    &  $16.7$    &  $17.2$    &  $-25.31$  &  $-25.44$      &            &  $5.18^\dagger$            &  {\scriptsize Axi. stellar Jeans }  &  {\scriptsize \citet{cappellarietal2013} } \\
NGC~4649   &  $16.5$    &  $16.5$    &  $15.7$    &  $-25.36$  &  $-25.48$      &  $4.73$    &  ${  7.96^*}$     &  {\scriptsize Axi. stellar orbit }  &  {\scriptsize \citet{ShenGebhardt2010} } \\
          &  $16.5$    &  $16.5$    &  $17.3$    &  $-25.36$  &  $-25.48$      &            &  $4.99^\dagger$            &  {\scriptsize Axi. stellar Jeans }  &  {\scriptsize \citet{cappellarietal2013} } \\
NGC~4889   &  $102.0$   &  $99.1$    &  $103.2$   &  $-26.64$  &  $-26.74$      &  $20.17$   &  ${  16.80^*}$    &  {\scriptsize Axi. stellar orbit }  &  {\scriptsize \citet{mcconnelletal2011a} } \\
NGC~5322   &  $34.2$    &  $31.5$    &  $30.3$    &  $-25.51$  &  $-25.43$      &            &  $3.54^\dagger$            &  {\scriptsize Axi. stellar Jeans }  &  {\scriptsize \citet{cappellarietal2013} } \\
NGC~5353   &  $41.1$    &  $34.8$    &  $35.2$    &  $-25.45$  &  $-25.31$      &            &  $3.15^\dagger$            &  {\scriptsize Axi. stellar Jeans }  &  {\scriptsize \citet{cappellarietal2013} } \\
NGC~5557   &  $51.0$    &  $49.2$    &  $38.8$    &  $-25.46$  &  $-25.68$      &            &  $2.73^\dagger$            &  {\scriptsize Axi. stellar Jeans }  &  {\scriptsize \citet{cappellarietal2013} } \\

NGC~7052   &  $69.3$    &  $61.9$    &  $58.7$    &  $-25.67$  &  $-25.63$      &            &  $3.06^\dagger$            &  {\scriptsize Sph. stellar Jeans }  &  {\scriptsize \citet{haringrix2004} } \\
   &  $69.3$    &  $61.9$    &  $58.7$    &  $-25.67$  &  $-25.63$      &  $0.35$    &                    &  {\scriptsize Ionized gas }  &  {\scriptsize \citetalias{vanderMarelvandenBosch1998} } 
\\
NGC~7619   &  $54.0$    &  $46.6$    &  $51.5$    &  $-25.65$  &  $-25.68$      &  $2.26$    &  ${ 4.98^*}$     &  {\scriptsize Axi. stellar orbit }  &  {\scriptsize \citet{ruslietal2013} } \\
\end{longtable*}
\endgroup

\section{The Stellar Mass Function}\label{appendix:phiL_phiH}
\setcounter{table}{0}
\renewcommand{\thetable}{\ref*{appendix:phiL_phiH}\arabic{table}}

\begin{table*}[t]
\centering
\begin{tabular}{lrrrrr}
\multicolumn{6}{c}{{\bf Table~\ref{tab:priorposterior}}}\\
\multicolumn{6}{c}{Posteriors for double-Schechter parameters for galaxy stellar mass functions}\\
\hline
 & \multicolumn{2}{c}{GSMF in \citet{lejaetal2020}} & \multicolumn{3}{c}{GSMF in this paper}  \\ 
  & $z = 0.2$ &  $z = 0.0$  & Dyn. $M_*$ & SPS $M_*$ & Combined $M_*$ \\ \hline
$\log_{10} (\phi_1 / \rm{Mpc}^{-3})$          & $-2.44\pm0.02$              & $-2.38\pm0.03$               & $-4.83\pm0.50$  & $-4.87\pm0.60$ & $-4.85\pm0.55$ \\
$\log_{10} (\phi_2 / \rm{Mpc}^{-3})$          & $-2.89^{+0.03}_{-0.04}$     & $-2.82\pm0.05$               & $-2.87\pm0.09$  & $-2.84\pm0.08$ & $-2.85\pm0.09$  \\
$\log_{10} (M_s/\msun)$             & $10.79\pm0.02$              & $10.77\pm0.03$               & $11.34\pm0.06$  & $11.31\pm0.06$ & $11.33\pm0.06$  \\
$\alpha_1$                  & $-0.28\pm0.07$              & $-0.28\pm0.07$               & $1.08\pm0.77$   & $0.84\pm0.96$  & $0.92\pm0.90$ \\
$\alpha_2$                  & $-1.48^{+0.01}_{-0.02}$     & $-1.48^{+0.01}_{-0.02}$      & $-1.38\pm0.04$  & $-1.37\pm0.04$ & $-1.38\pm0.04$  \\
\hline
$\log_{10} (\phi_\textrm{L}/ \rm{Mpc}^{-3})$ & $-2.10\pm0.04$ & $-2.04\pm0.04$ &  $-2.01\pm0.03$ &  $-2.01\pm0.03$ & $-2.01\pm0.03$ \\
$\log_{10} (\phi_\textrm{H}/ \rm{Mpc}^{-3})$ & $-8.25\pm0.32$ & $-8.57\pm0.43$ &  $-4.45\pm0.09$ &  $-4.59\pm0.11$ & $-4.52\pm0.12$ \\
\end{tabular}
\caption{
Posterior distributions for the parameters of double-Schechter approximation to five GSMFs. The bottom two rows list GSMF amplitudes at two reference masses, $\phi_\textrm{L}$ at $M_\textrm{L} = 10^{10}\msun$ and $\phi_\textrm{H}$ at $M_\textrm{H} = 10^{12}\msun$.  Column 1: parameter names. Column 2: $z=0.2$ GSMF of \citet{lejaetal2020}. Column 3: Our extrapolation of \citet{lejaetal2020} to $z=0$.  Column 4: GSMF of this paper constructed from the lower-mass part of the $z=0$ GSMF of \citet{lejaetal2020} and the MASSIVE GSMF at higher masses, using dynamically inferred $M_*$ for MASSIVE galaxies.  Column 5: same as Column 4 but for MASSIVE SPS inferred $M_*$.  Column 6: the fiducial GSMF of this paper, combining the two posteriors of Columns 4 and 5. All errors denote 68\% confidence intervals.  } 
\label{tab:priorposterior}
\end{table*}

The GSMF in this work has a double Schechter form (Equation~\ref{eq:gsmf}). To determine the parameters in this function, we use the procedure of Section~4.2 of \citet{lejaetal2020} and model the MASSIVE observations as the result of an inhomogeneous Poisson process whose rate function is a product of the GSMF and the differential comoving volume. As discussed in Section~\ref{sec:MstarDist}, when using masses inferred from \mknew\ we use distances which have been homogenized with the SBF measurements to compute the differential comoving volumes. We add a $\chi^2$-like term to the log-likelihood that is associated with the mis-match between the proposed GSMF and that of \citet{lejaetal2020} at three stellar masses $10^9$, $10^{10}$, and $10^{11}\msun$. This choice of points approximately recovers the 90\% error band of their GSMF between $10^9\msun$ and $10^{11}\msun$ without significantly degrading the fit to the MASSIVE data at high masses.

We use \texttt{dynesty} \citep{Speagleetal2020DYNESTY} to sample the posterior distribution for the five model parameters $\phi_1, \phi_2, \alpha_1, \alpha_2$ and $M_s$. The posterior distribution exhibits a strong correlation between the amplitude and power-law slope of the high-mass component, $\phi_1$ and $\alpha_1$. This correlation is due to the fact that the high-stellar-mass observations in MASSIVE lie well above the preferred scale radius $M_s$, where the shape of the GSMF is largely set by the exponential term and where changes in the power-law slope result primarily in a change in the amplitude of the component rather than in a change to the shape of the function. Accordingly, the power-law slope and the amplitude of the high-mass component are correlated in our posterior distribution.

Because of this, approximating the posterior on these parameters as uncorrelated will significantly over-represent the true uncertainty in the GSMF. However, when $\phi_1$ and $\phi_2$ are replaced with the amplitudes of the GSMF at two widely separated mass scales, e.g., $M_\textrm{L} = 10^{10}\msun$ and $M_\textrm{H} = 10^{12}\msun$, we find the posterior to be reasonably well approximated with uncorrelated Gaussians.  We perform the posterior sampling using the standard parameters $(\phi_1,\phi_2,M_s,\alpha_1,\alpha_2)$ and present the inferred values on these parameters and the values of the GSMF at $10^{10}\msun$ and $10^{12}\msun$ ($\phi_\textrm{L}$ and $\phi_\textrm{H}$) in Table~\ref{tab:priorposterior}. 

When reproducing the posterior distribution of our GSMF fit parameters, we recommend drawing uncorrelated Gaussian distributions for $(\phi_\textrm{L},\phi_\textrm{H},M_s,\alpha_1,\alpha_2)$ with the centers and widths listed in Table~\ref{tab:priorposterior}, and then using the following script to map these parameters into the standard $\phi_1$ and $\phi_2$ that appear in Equation~(\ref{eq:gsmf}). For comparison, when the full posterior distribution is used, the 90\% errorband of the GSMF for dynamical $M_*$ is on average 0.39 dex wide between $10^{11.5}\msun$ and $10^{12.5}\msun$, while it is 0.47 dex wide when using the scheme described below, and is inflated to 1.16 dex wide when naively drawing $\alpha_1$ and $\phi_1$ as uncorrelated variables.

Finally, we combine the posteriors when inferring stellar masses from SPS-based measurements and dynamical measurements. This is trivially done by drawing one half of the realizations of the combined posterior from the SPS-based posterior and one half from the dynamical-$M_*$-based posterior.

For ease of reproduction, we have provided below a short section of \texttt{python} code which can be used to approximate the posteriors of the parameters of our GSMF.

\noindent\rule{\columnwidth}{1pt}

\definecolor{backcolour}{rgb}{0.98,0.98,0.98}
\begin{lstlisting}[language=Python,backgroundcolor=\color{backcolour},  basicstyle=\ttfamily\scriptsize,
  breakatwhitespace=false,         
  breaklines=true,                 
  captionpos=b,                    
  keepspaces=true,                 
  %numbers=left,                    
  numbersep=5pt,                  
  showspaces=false,                
  showstringspaces=false,
  showtabs=false,                  
  tabsize=2
]

import numpy as np

def compute_phi1_phi2(
    phi_L,
    phi_H,
    alph_1,
    alph_2,
    log_Ms,
    log_ML = 10,
    log_MH = 12):

    tL = 10**(log_ML - log_Ms)
    tH = 10**(log_MH - log_Ms)

    l_1L = np.exp(-tL)*tL**(alph_1+1)
    l_1H = np.exp(-tH)*tH**(alph_1+1)
    l_2L = np.exp(-tL)*tL**(alph_2+1)
    l_2H = np.exp(-tH)*tH**(alph_2+1)

    denominator = np.log(10) * \
        (l_1L * l_2H - l_1H * l_2L)

    phi1 = l_2H * phi_L - l_2L * phi_H /\
        denominator
    phi2 = l_1L * phi_H - l_1H * phi_L /\
        denominator

    phi1[phi1 < 0] = 1e-10
    phi2[phi2 < 0] = 1e-10

    return(phi1, phi2)

n_post_realizations = 1000000
n_post_dyn_realizations = n_post_realizations // 2
n_post_sps_realizations = n_post_realizations // 2

# These values are from Table B1

post_dyn_values = np.array([
    [-2.01, 0.03],
    [-4.52, 0.12],
    [ 0.92, 0.90],
    [-1.38, 0.04],
    [11.33, 0.06],
    ])

post_sps_values = np.array([
    [-2.01, 0.03],
    [-4.59, 0.11],
    [ 0.84, 0.96],
    [-1.37, 0.04],
    [11.31, 0.06],
    ])

post_dyn_realizations = np.random.normal(
    post_dyn_values[:,0],
    post_dyn_values[:,1],
    size=(n_post_dyn_realizations,5))

post_sps_realizations = np.random.normal(
    post_sps_values[:,0],
    post_sps_values[:,1],
    size=(n_post_sps_realizations,5))

posterior_realizations = np.r_[
    post_dyn_realizations,post_sps_realizations
    ]

posterior_realizations[:,:2] = \
    10**posterior_realizations[:,:2] 

posterior_realizations[:,0], \
    posterior_realizations[:,1] = \
    compute_phi1_phi2(*posterior_realizations.T)

posterior_realizations[:,:2] = \
    np.log10(posterior_realizations[:,:2])



\end{lstlisting}

\noindent\rule{\columnwidth}{1pt}

\bibliography{gsmf}{}
\bibliographystyle{aasjournal}

\end{document}